\documentclass[11pt,oneside,onecolumn,fleqn,a4paper]{article}

\usepackage{amssymb,amsmath,pstricks}
\allowdisplaybreaks[1]

\usepackage{amsfonts}

\let\epsilon=\varepsilon

\def\Proof{\noindent{\bf Proof}\quad}

\def\R{\mathcal{R}}

\def\QED{\nolinebreak \hspace*{\fill} \ensuremath{\square}}

\def\restrict{\! \upharpoonright \!}

\def\vrT{\mathrm{VR\!-\!totality}}
\def\vrLT{\mathrm{VR(log)\!-\!totality}}
\def\vrT2{\mathrm{2VR\!-\!totality}}
\def\vrLT2{\mathrm{2VR(log)\!-\!totality}}

\newcommand{\ignore}[1]{}

\begin{document}

\newtheorem{theo}{Theorem} % [section]
\newtheorem{prop}[theo]{Proposition}
\newtheorem{lemma}[theo]{Lemma}
\newtheorem{cor}[theo]{Corollary}
\newtheorem{rem}[theo]{Remark}
\newtheorem{con}[theo]{Conjecture}
\newtheorem{claim}[theo]{Claim}
\newtheorem{defn}[theo]{Definition}

\title{Notes on switching lemmas\thanks{
These notes were originally published online in May 2009 at 
\texttt{users.math.cas.cz/$\sim$thapen/switching.pdf}.
}}
\author{Neil Thapen\thanks{
Institute of Mathematics, Academy of Sciences of the Czech Republic, \texttt{thapen@math.cas.cz}. 
Partially supported by Institutional Research Plan AV0Z10190503 and grant IAA100190902 of GA AV \v{C}R, 
and by a grant from the John Templeton Foundation.}}
\maketitle

We prove three switching lemmas, for random restrictions for which variables are set independently;
for random restrictions where variables are set in blocks (both due to H{\aa}stad \cite{hastad}); 
and for a distribution appropriate
for the bijective pigeonhole principle \cite{bip, kpw}. The proofs are based on Beame's version 
\cite{beame} of Razborov's proof of the switching lemma in \cite{razborov}, 
except using families of weighted restrictions 
rather than families of restrictions which are all the same size. 
This follows a suggestion of Beame in \cite{beame}.
The result is something between H{\aa}stad's and Razborov's methods of proof.
We use probabilistic arguments rather than counting ones, in a similar
way to H{\aa}stad, but rather than
doing induction on the terms in our formula with an inductive
hypothesis involving conditional probability, as H{\aa}stad does, we 
explicitly build one function to bound the probabilities for the whole formula.

\section{A restriction which sets variables independently}

Let $F$ be an $r$-DNF, that is, a disjunction of conjunctions (which we will usually
call ``terms'') where each conjunction has size $r$ or less.
Suppose the variables in $F$ come from a set $X$ of size $n$.

Fix a probability $p$. Define a distribution $\R$ of partial restrictions to the variables
by choosing a restriction $\rho$ as follows: independently for each $x \in X$, set $x$ to $0$
with probability $\frac{1-p}{2}$; to $1$ with probability $\frac{1-p}{2}$; 
or to $*$ (meaning ``leave it unset'') with probability $p$. 

The weight $|\rho|$ of a restriction $\rho \in \R$ is its probability of being chosen from $\R$. 
So if $\rho$ has exactly $a$ many $1$s, $b$ many $0$s and $c$ many $*$s, then the weight
of $\rho$ is $(\frac{1-p}{2})^{a+b}p^c$. The weight $|S|$ of a set of restrictions 
$S \subseteq \R$ is the probability that a random restriction from $\R$ belongs to $S$, or, 
equivalently, the sum of the weights of the restrictions in $S$.

The canonical tree $T(F,\rho)$ is defined by the following decision procedure:
Look through $F$ for a term $C$ such that $C \restrict \rho \not \equiv 0$.
If there is no such term, then halt and output ``0''. Otherwise let
$C_1$ be the first such term. Let $\beta_1$ list the variables that appear starred 
 in $C_1 \restrict \rho$.
Query all these variables in order, and let the assignment $\pi_1$ be given by the replies. 
If $\rho \pi_1$ satisfies $C_1$ (or if $\beta_1$ was
empty and $C_1$ was already satisfied in $\rho$) halt and output ``1''.
Otherwise repeat this step starting with $\rho \pi_1$ in place of $\rho$, 
looking for a term $C_2$ which is the first such that $C_2 \restrict \rho \pi_1 \not \equiv 0$
etc., until we run out of terms.

Note that this tree correctly decides $F \restrict \rho$, in that if $\pi$ is given by the answers
along a branch in the tree, then the
label at the end of the branch (called the ``output'' above) is the value of $F \restrict \rho \pi$.

\begin{lemma}
Fix a number $s>0$. Let $S$ be the set of restrictions $\rho$ in $\R$ for which 
$T(F,\rho)$ has height $s$ or greater. Then $|S| \le (9pr)^s$ (assuming $p<1/9$).
\end{lemma}

\Proof
We will bound the weight of $S$ by defining an injection from $S$ into a set of small
weight (roughly speaking) and then arguing about how this map changes weights.
So let $\rho \in S$ and let $\pi$ be the first path in $T(F, \rho)$ with length 
$s$ or greater. 

Let $C_1, \ldots, C_k$, $\beta_1, \dots, \beta_k$ and $\pi_1, \dots, \pi_k$
be the terms, unset variables and assignments to them encountered along $\pi$,
as far as the $s$th query in $\pi$, from the construction
of $T(F, \rho)$. It may be that the $s$th query in $\pi$ occurred in the middle of
querying the variables $\beta_k$, in which case we trim $\beta_k$ and $\pi_k$ to only
include the variables mentioned in the first $s$ queries in $\pi$.

For each $i=1, \dots, k$ let $\sigma_i$ be the (unique) assignment to $\beta_i$ 
which is consistent with $C_i \restrict \rho \pi_1 \dots \pi_{i-1}$
(for $i<k$, this will actually satisfy $C_i \restrict \rho \pi_1 \dots \pi_{i-1}$).
Note that the $\beta_i$s
are all disjoint so that $\rho \sigma_1 \dots \sigma_k$ is a well-defined restriction.
Let $\sigma$ be $\sigma_1 \dots \sigma_k$.

We will code each $\beta_i$ as a string $\beta'_i$ of $|\beta_i|$ numbers, each less than $2r$,
by recording, for each variable in $\beta_i$, its location in $C_i$ (a number less than $r$)
and whether it is the last variable in $\beta_i$ (one bit). 
We will code the whole tuple $\beta_1, \dots, \beta_k$ as the concatenation 
$\beta' = \beta'_1 \dots \beta'_k$ of these strings. So over $S$ there are $(2r)^s$ possible different
strings $\beta'$.

We will code each $\pi_i$ simply as a string $\pi'_i$ of $|\beta_i|$ bits, one for each variable in $\beta_i$.
$\pi'$ will be the concatenation $\pi'_1 \dots \pi'_k$. So there are $2^s$ possible strings
$\pi'$.

We now define a map $\theta : S \rightarrow \R \times (2r)^s \times 2^s$
by \[
\theta : \rho \mapsto (\rho \sigma, \beta', \pi').
\]

We claim that this is an injection. To see this, suppose we are given $(\rho \sigma, \beta', \pi')$.
We may recover $\rho$ as follows:
First, we can easily recover all strings $\beta'_i$ and $\pi'_i$.
Now let $C'_1$ be the first term in $F$ such that 
$C'_1 \restrict \rho \sigma \not \equiv 0$. $C'_1$ cannot come before $C_1$, and by construction
$C_1$ is not falsified by $\rho \sigma$. So we must have $C'_1 = C_1$. From $\beta'_1$
and $C_1$ we can recover $\beta_1$, and from this and $\pi'_1$ we can recover $\pi_1$. 
$\sigma_1$ and $\pi_1$ were assignments to the same variables,
so we can construct a restriction $\rho \sigma [\pi_1 / \sigma_1] = \rho \pi_1 \sigma_2 \dots \sigma_k$.
Let $C'_2$ be the first term in $F$ such that $C'_2 \restrict \rho \pi_1 \sigma_2 \dots \sigma_k \not \equiv 0$.
Then as above, $C'_2$ must equal $C_2$, and we can recover $\beta_2$ and $\pi_2$ and
carry on in the same way.
Once we have recovered all the $\beta_i$s we know exactly what changed between $\rho$ and
$\rho \sigma$ and can recover $\rho$. 

Now temporarily fix some values of $\beta'$ and $\pi'$ and let $S_{\beta',\pi'}$
be the subset of $S$ consisting of all $\rho$s to which $\theta$ assigns these values.

Restricted to $S_{\beta',\pi'}$, the first component $\theta_1 : \rho \mapsto \rho \sigma$
of $\theta$ is an injection $S_{\beta',\pi'} \rightarrow \R$, so the weight $|\theta_1[S_{\beta',\pi'}]|$
of its image is the sum of the individual weights 
$|\theta_1(\rho)|$ over all $\rho \in S_{\beta',\pi'}$. But $\rho \sigma$ sets exactly $s$ variables that
were unset in $\rho$, so $|\rho \sigma| = p^{-s} (\frac{1-p}{2})^s  |\rho|$.
Hence 
\[
|\theta_1[S_{\beta',\pi'}]| = \left(\frac{1-p}{2p}\right)^s |S_{\beta',\pi'}|.
\]
But $\theta_1[S_{\beta',\pi'}]$ is a subset of $\R$ so has weight $\le 1$, 
so $|S_{\beta',\pi'}| \le (\frac{2p}{1-p})^s$.

Finally $S$ is the union of the sets $S_{\beta',\pi'}$ over all possible
strings $\beta',\pi'$. So
\[
|S| \le (2r)^s 2^s \left(\frac{2p}{1-p}\right)^s = \left(\frac{8pr}{1-p}\right)^s
\]
giving the result. 
\QED

\section{A restriction which sets variables in blocks}

Let $F$ be an $r$-DNF in variables $X$, as above.
Suppose that $X$ is partitioned into a family $\mathcal{B}$ of disjoint blocks. 
We assume there is some fixed ordering on the variables in each block.

Fix probabilities $p$ and $q$ (in the usual application we may take $p=q$, but it is useful
to keep them separate to keep track of what is happening in the proof). 
Define a distribution $\R$ of partial restrictions 
by choosing a restriction $\rho$ in two stages, as follows: First, 
independently for each $x \in X$, set $x$ to $1$
with probability $1-p$, otherwise leaving it starred.
Then, independently for each block $B \in \mathcal{B}$ (ignoring any
blocks which are already set to all $1$), 
with probability $1-q$ set all starred variables in $B$ to $0$ (in which case we call $B$ a $0$-block),
otherwise leaving them all starred (in which case we call $B$ a $*$-block).

The weight of a restriction $\rho$ is the product of the weights 
of its blocks. If a block has $a$ many $1$s and $b>0$ many non-$1$s,
its weight is $(1-p)^ap^b(1-q)$ if it is a $0$-block and $(1-p)^ap^bq$
if it is a $*$-block. If a block is all $1$s (in which case the terms
$0$-block and $*$-block become meaningless) then its weight is $(1-p)^{|B|}$ 
(although in applications there are unlikely to be
any such blocks).

The restriction $g(\rho)$ extends $\rho$ further: for each $*$-block in $\rho$, 
it sets every starred variable, except for the first one, to be $1$.

The canonical tree $T(F,\rho)$ is defined by the following decision procedure:
Look through $F$ for a term $C$ such that $C \restrict \rho \not \equiv 0$.
If there is no such term, then halt and output ``0''. Otherwise let
$C_1$ be the first such term. Let $\beta_1$ list the blocks
$B$ such that a starred variable from $B$ appears in
$C_1 \restrict \rho$, in the order in which they appear.
For $B \in \beta_1$, query the single starred variable in $B \restrict g(\rho)$,
regardless of whether or not this is the variable appearing in $C_1 \restrict \rho$.
Let $\pi_1$ be the complete assignment to all the blocks in $\beta_1$ given by $g(\rho)$
together with all the replies to the queries. That is, under $\rho \pi_1$ 
each block $B \in \beta_1$ will be set to $1$
everywhere, except that the variable that was queried may be set to either $0$ or $1$, 
depending on the reply.
If $\rho \pi_1$ satisfies $C_1$ (or if $C_1$ was already satisfied in $\rho$) halt and output ``1''.
Otherwise repeat this step starting with $\rho \pi_1$ in place of $\rho$, 
etc.

Note that along any branch in the tree no block is queried more than once, and that the
moment a block is queried all variables in that block are given a $0$ or $1$ value.

The tree $T(F,\rho)$ correctly decides $F \restrict g(\rho)$, in that if $\pi$ is given by the answers along
a branch in the tree, then the
label at the end of the branch is the value of $F \restrict g(\rho) \pi$.

\begin{lemma}
Fix a number $s>0$. Let $S$ be the set of restrictions $\rho$ in $\R$ for which 
$T(F,\rho)$ has height $s$ or greater. Then $|S| \le (13qr)^s$ (assuming $p<1/2r$ and $q<1/13$).
\end{lemma}

\Proof
The proof is  similar to the proof of the previous lemma.
Let $\rho \in S$ and let $\pi$ be the first path in $T(F, \rho)$ with length 
$s$ or greater. 

Let $C_1, \ldots, C_k$, $\beta_1, \dots, \beta_k$ and $\pi_1, \dots, \pi_k$
be the terms, blocks and assignments encountered along $\pi$,
as far as the $s$th query in $\pi$, from the construction
of $T(F, \rho)$. If necessary, trim $\beta_k$ and $\pi_k$ to only
include the blocks mentioned in the first $s$ queries in $\pi$.

For $i= 1, \dots, k$ let $\gamma_i$ list the starred variables from the
blocks in $\beta_i$ which appear positively in $C_i \restrict \rho \pi_1 \dots \pi_{i-1}$.

For $i=1, \dots, k$ let $\sigma_i$ be the following assignment. For each block 
$B \in \beta_i$, $\sigma_i$ sets every starred (under $\rho$)
variable in $B$ that appears positively in 
$C_i \restrict \pi_1 \dots \pi_{i-1}$ to $1$, and sets all remaining starred variables in $B$ to $0$.
Note that this is consistent with $C_i \restrict \pi_1 \dots \pi_{i-1}$, and also that
$\sigma_i$ sets exactly the same variables as $\pi_i$ does, so that
the different $\sigma_i$s are disjoint and $\rho \sigma_1 \dots \sigma_k$ is a well-defined restriction.
Let $\sigma$ be $\sigma_1 \dots \sigma_k$.

As before, we will code $\beta_1, \dots, \beta_k$ as a string $\beta' = \beta'_1 \dots \beta'_k$,
by recording the location of the first starred variable from a block. There are $(2r)^s$
 possible strings $\beta'$.
 
As before, we will code $\pi_1, \dots, \pi_k$ as a string
$\pi' = \pi'_1 \dots \pi'_k$ of $s$ bits.

We will code each $\gamma_i$ as a string $\gamma'_i$ of $r$ bits, one for each literal
in $C_i$, recording whether that literal is in $\gamma_i$. We code the whole tuple
$\gamma_1, \dots, \gamma_k$ as the concatenation $\gamma'=\gamma'_1 \dots \gamma'_k$ of
these strings. There are at most $2^{rs}$ many possible $\gamma'$s. 

We define $\theta : S \rightarrow \R \times (2r)^s \times 2^s \times 2^{rs}$
by \[
\theta : \rho \mapsto (\rho \sigma, \beta', \pi', \gamma').
\]

To see that this is an injection, suppose we are given $(\rho \sigma, \beta', \pi', \gamma')$.
We can recover $C_1$ and $\beta_1$ just as in the previous lemma, and from $C_1$
and $\gamma'_1$ we can recover $\gamma_1$. But now for each block $B \in \beta_1$,
$\gamma_1$ tells us exactly what we changed to go from $\rho \restrict B$ to 
$\rho \sigma_1 \restrict B$. We can undo the changes, and recover $\rho \restrict B$, 
by setting all variables in $B$ mentioned in $\gamma_1$ to $*$ and setting all $0$s
in $\rho \sigma_1 \restrict B$ to $*$. Then we can recover $\pi_1 \restrict B$ by setting
all but the first $*$ to $1$ and setting the first $*$ according to $\pi'_1$. 
Then we continue for the rest of the terms, similarly to the previous lemma.

Now temporarily fix some values of $\beta'$, $\pi'$ and $\gamma'$ and let $S_{\beta',\pi',\gamma'}$
be the subset of $S$ consisting of all $\rho$s to which $\theta$ assigns these values.
Let $\theta_1$ be as before, and suppose that $\gamma'$ records $m$ many variables in total.

For $\rho \in S_{\beta',\pi',\gamma'}$, 
going from $\rho$ to $\rho \sigma$ changes $m$ many non-$1$s into $1$s, which increases the weight
by a factor of $(\frac{1-p}{p})^m$. It also changes $s$ many $*$-blocks, each one being 
changed into either a $0$-block, increasing the weight by a factor of $\frac{1-q}{q}$,
or into an all-$1$ block, increasing the weight by a factor of $\frac{1}{q}$. 
So the total increase is $\ge(\frac{1-p}{p})^m (\frac{1-q}{q})^s$. 
Hence
\[
|S_{\beta',\pi',\gamma'}| \le \left(\frac{p}{1-p}\right)^m \left(\frac{q}{1-q}\right)^s.
\]
For $\rho \in S$ there are at most $rs$ variables that could be recorded in $\gamma'$.
So we can calculate
\begin{align*}
|S_{\beta',\pi'}| = \sum_{\gamma'} |S_{\beta',\pi',\gamma'}| 
&= \sum_{m=0}^{rs} \sum_{|\gamma'|=m} |S_{\beta',\pi',\gamma'}| \\
&\le \sum_{m=0}^{rs} \binom{rs}{m}  \left(\frac{p}{1-p}\right)^m \left(\frac{q}{1-q}\right)^s \\
&= \left( 1+\frac{p}{1-p} \right)^{rs} \left(\frac{q}{1-q}\right)^s \\
&\le e^\frac{prs}{1-p} \left(\frac{q}{1-q}\right)^s \\
&\le \left(\frac{3q}{1-q}\right)^s
\end{align*}
where the last inequality holds if $p<\frac{1}{2r}$.

Finally 
\[
|S| \le (2r)^s 2^s \left(\frac{3q}{1-q}\right)^s = \left(\frac{12qr}{1-q}\right)^s
\]
giving the result. Clearly the constant could be improved by putting some more
conditions on $p$ and $q$. 
\QED

\section{A restriction for the pigeonhole principle}

Let $F$ be an $r$-DNF in variables $P = \{ p_{xy} : x \in n+1, y \in n \}$,
where we take $p_{xy}$ to express that ``pigeon $x$ goes to hole $y$''.

Fix a probability $q$. Define a distribution $\R$ of partial injections of 
$n+1$ into $n$ by choosing a partial injection $\rho$ as follows: choose
the range of $\rho$ by putting each hole into the range independently 
at random with probability $(1-q)$, then choose uniformly at random
from all possible injections from the set of pigeons into this range
(we do this ``backwards'' to exclude the possibility of having to find holes for $n+1$ pigeons).
If $\rho$ sets exactly $m$ pigeons, then the weight of $\rho$ is 
$(1-q)^m q^{n+1-m} \frac{(n+1-m)!}{(n+1)!}$.

We will identify a partial injection $\rho$ with the partial assignment
to the variables $P$ in which, for each pigeon $x$ which is sent to a hole $y$
by $\rho$: $p_{xy}$ is set to true, $p_{xy'}$ is set to false for each $y' \neq y$,
and $p_{x'y}$ is set to false for each $x' \neq x$.

To define the canonical tree $T(F,\rho)$ for $F$ with respect to such a $\rho$,
we first pre-process $F$.
For each term in $F$ we replace each negative literal
$\neg p_{xy}$ with the disjunction $\bigvee_{y' \neq y} p_{xy}$ and then 
distribute out so that the formula is once again an $r$-DNF (which may now possibly be
$n^r$ times larger). We then remove any term which asserts that two 
pigeons go to one hole or that one pigeon goes to two holes.
We call our new $r$-DNF $F'$.

The tree will again be given by a decision procedure. This time
the queries are not to the values of propositional variables. 
Instead we can either name a pigeon and query which hole it
goes to, or name a hole and query which pigeon goes to it. 
We may assume that the replies always form a partial
injection (assume that the tree halts in some error state
if a branch becomes so long that there are no free holes or
pigeons available to reply to a query).

Now look for the first term $C$ in $F'$ such that $C \restrict \rho \neq 0$.
If there is no such term, halt and output ``0''.
Otherwise let $C_1$ be the first such term. Let $\beta_1$ list the literals
$p_{xy}$ that appear in $C_1 \restrict \rho$. For each such $p_{xy}$, query
which hole pigeon $x$ goes to, and then query which pigeon goes to hole $y$.
Let $\pi_1$ be the partial injection given by the replies to these queries.
As before, if $\rho \pi_1$ satisfies $C_1$ halt and output ``1'', and
otherwise repeat this step starting with $\rho \pi_1$ in place of $\rho$, 
etc.

Let $\pi$ be a branch in the tree, which we will identify with the partial
injection given by the replies along that branch. If $\pi$ ends with
the output ``1'' at the leaf, then by construction $\pi$ must satisfy
some term $C'$ in $F'$ and hence must also satisfy the term $C$ in $F$
from which $C$ arose. Conversely, if $\pi$ ends with the output ``0'',
then there is no extension of $\pi$ to a partial injection $\alpha$ 
which satisfies any term $C$ in $F$. For suppose such an $\alpha$ and $C$
existed. Then $\alpha$ would also satisfy some term $C'$ in $F'$ arising from $C$;
but then $C' \restrict \rho$ cannot be $0$, contradicting the construction of the tree.

\begin{lemma}
Fix a number $s>0$. Let the probability $q$ be chosen so that $128 r^2 n^3 q^4 < 1$
(the constant here could easily be improved).
Let $S$ be the set of partial injections $\rho \in \R$
for which $T(F,\rho)$ has height $s$ or greater. Then the weight $|S|$ is
exponentially small in $s$.
\end{lemma}

\Proof
Let $l = 2qn$. By the Chernoff bound, for all but an exponentially
small (in $n$) number of exceptions, every $\rho \in S$ leaves fewer
than $l$ pigeons and $l$ holes unset. Hence we may assume in what follows
that every $\rho \in S$ has this property.

Let $\rho \in S$ and let $\pi$ be the first path in $T(F, \rho)$ with length 
$s$ or greater. 

Let $C_1, \ldots, C_k$, $\beta_1, \dots, \beta_k$ and $\pi_1, \dots, \pi_k$
be the terms, literals and replies encountered along $\pi$,
as far as the $s$th query in $\pi$, from the construction
of $T(F, \rho)$. If necessary, trim $\beta_k$ and $\pi_k$ to only
include the blocks mentioned in the first $s$ queries in $\pi$.

For each $i=1, \ldots, k$ let $\sigma_i$ be the partial injection 
which is specified by the literals in $\beta_i$.
Note that every pigeon and hole in $\sigma_i$ also occurs in $\pi_i$,
because in the tree we query both the pigeons and the holes occurring in $\beta_i$.
Hence for $j>i$, $\sigma_i$ is consistent with $\sigma_j$,
because the pigeons and holes occurring in $\sigma_j$ are disjoint 
from those in $\pi_i$ and hence disjoint from those in $\sigma_i$
(recall that $\beta_j$ lists the literals $p_{xy}$ in $C_j \restrict \rho \pi_1 \ldots \pi_{j-1}$).
Let $\sigma$ be $\sigma_1 \dots \sigma_k$.

As before, we will code $\beta_1, \dots, \beta_k$ as a string
$\beta' = \beta'_1 \dots \beta'_k$, by recording the locations
of the literals $\beta_i$ in each term $C_i$. There are $(2r)^s$
possible strings $\beta'$.
 
Let $A$ and $B$ be respectively the set of pigeons and the
set of holes left unset in $\rho \sigma$. We know that $|A|, |B| \le l$.
We will code $\pi_i$ as follows. For each $p_{xy}$ in $\beta_i$, 
pigeon $x$ was queried in the branch $\pi$ and was either assigned to hole $y$
or to some hole $y' \neq y$. In the second case, such a $y'$ must be in $B$ because from 
this point onwards, neither
the tree nor $\sigma$ can assign any other pigeon to hole $y'$. 
Hence we may code which hole was assigned using a single bit and a number less than $l$. 
Similarly hole $y$ was assigned either pigeon $x$ or some pigeon from $A$,
and we code this in a similar way. Let $\pi'_i$ be the string coding all replies in
$\pi_i$ in this way, and let $\pi'$ be the concatenation $\pi'_1 \dots \pi'_k$. 
There are $(2l)^s$ possible strings $\pi'$.

We define $\theta : S \rightarrow \R \times (2r)^s \times (2l)^s$ by
\[
\theta : \rho \mapsto (\rho \sigma, \beta', \pi').
\]
This is an injection, for suppose we are given $(\rho \sigma, \beta', \pi')$.
We know $A$ and $B$ immediately. We can recover $C_1$ and $\beta_1$ as before,
and from $\beta_1$, $A$ and $B$ we can recover $\pi_1$. Then we continue as before.

Now temporarily fix some values of $\beta'$ and $\pi'$ and let $S_{\beta',\pi'}$
be the subset of $S$ to which $\theta$ assigns these values. Let $\theta_1$
be as before. 

For $\rho \in S_{\beta',\pi'}$, going from $\rho$ to $\rho \sigma$ sets at least
$s/2$ more pigeons. Adding one more pigeon to a partial injection of size $m$ changes its
weight by a factor of $\frac{1-q}{q}\frac{1}{n+1-m}$. In our case $n-m$ is always smaller than 
$l$, so the factor is at least $\frac{1-q}{ql}$. Hence the total increase is 
at least $(\frac{1-q}{ql})^{s/2}$, and thus
$|S_{\beta',\pi'}| \le ( \frac{ql}{1-q} )^{s/2}$ which
we can bound by $(2ql)^{s/2}$ if we assume that $q<1/2$.

Finally, recalling that $l=2qn$, we have
\begin{align*}
|S| \le (2r)^s (2l)^s (2ql)^{s/2} 
&= (4 r^2 \cdot 16 q^2 n^2 \cdot 2 q^2 n )^{s/2}\\
&= (128 r^2 n^3 q^4)^{s/2},
\end{align*}
which gives the result.
\QED

\end{document}